\documentclass[10pt]{iopart}
\usepackage{iopams,amssymb,bm,verbatim,hyperref}
\usepackage[T1]{fontenc}
\usepackage[perpage]{footmisc} 
\usepackage{xcolor} 
\newcommand{\smfrac}[2]{\textstyle{\frac{#1}{#2}}}
\newcommand{\half}{\smfrac{1}{2}}
\newcommand{\be}{\begin{equation}}
\newcommand{\ee}{\end{equation}}
\newcommand{\w}[1]{\bm{#1}}
\newcommand{\Journal}[4]{{#4} {\it #1} {\bf #2}, #3 }
\newcommand{\tho}{\textrm{\TH}}
\newcommand{\thd}{\tho '}
\newcommand{\et}{\eth}
\newcommand{\etd}{\eth '}
\newcommand{\ud}{\textrm{d}}
\newcommand{\kc}{\overline{\kappa}}
\newcommand{\lc}{\overline{\lambda}}
\newcommand{\muc}{\overline{\mu}}

\newcommand{\nuc}{\overline{\nu}}
\newcommand{\pic}{\overline{\pi}}
\newcommand{\rhoc}{\overline{\rho}}

\newcommand{\sigmac}{\overline{\sigma}}
\newcommand{\Psic}{\overline{\Psi}}
\newcommand{\tauc}{\overline{\tau}}
\def\f{f}
\def\g{g}
\def\h{h}
\def\j{j}
\def\x{t}
\def\y{z}
\def\nx{x}
\def\ns{s}
\def\u{u} 
\def\v{v} 
\def\Eu{\mathcal{R}}
\def\Ev{\mathcal{S}}
\def\Ew{w}
\def\EW{w'}
\def\EWc{\overline{\EW}}
\def\Ewc{\overline{\Ew}}
\def\EJ{\mathcal{T}}
\def\Czero{C_0}
\def\nB{\mathcal{B}}
\def\nC{\mathcal{C}}
\def\nP{\mathcal{P}}
\def\nQ{\mathcal{Q}}
\def\nvarsigma{\varsigma}
\def\nxi{\xi}
\def\nF{F}
\def\nJ{J}
\def\nH{H}
\def\nK{q}
\def\nN{N}
\def\nNu{\nN_{,\u}}
\def\nNv{\nN_{,\v}}
\def\nNuu{\nN_{,\u\u}}
\def\nNvv{\nN_{,\v\v}}
\def\nCHI{\Xi}
\def\nSigma{\Sigma}

\def\wzero{w_0}
\def\psi{\Psi_2}
\def\nO{\w{\Omega}}
\def\np{p}
\def\npp{p}
\def\nq{q}
\def\zero{\Phi_0}
\def\zeroc{\overline{\Phi_0}}
\def\one{\Phi_1}
\def\onec{\overline{\Phi_1}}
\def\two{\Phi_2}
\def\twoc{\overline{\Phi_2}}

\begin{document}
\title{Non-aligned Einstein-Maxwell Robinson-Trautman fields of Petrov type D}
\author{Norbert Van den Bergh$^1$, John Carminati$^2$}
\address{$^1$\ Ghent University, Dept.~of Electronics and Information Systems, Ghent, Belgium}
\address{$^2$\ School of Information Technology, Deakin University, Australia}
\eads{\mailto{norbert.vandenbergh@gmail.com}, \mailto{jcarm1930@gmail.com}}

\begin{abstract}We discuss Petrov type D Einstein-Maxwell fields in which both double null eigenvectors of the Weyl tensor are non-aligned with the eigenvectors of a non-null electromagnetic field and
are assumed to be geodesic, shear-free, diverging and non-twisting. We obtain the general solution of the Einstein-Maxwell equations under the extra condition that the complex null vectors of the Weyl canonical tetrad are 
hypersurface orthogonal. The corresponding space-times are all conformally related to a Killing-Yano space and are described by a 5-parameter
family of metrics, admitting two commuting Killing vectors and having the C-metric as a possible vacuum limit.
\end{abstract}

\section{Introduction}
In the quest for exact solutions of the Einstein-Maxwell (EM) equations considerable research has been devoted to the study of aligned
EM fields, in which at least one of the principal null directions (PNDs) of the electromagnetic field $\mathbf {F}$ is parallel to a PND of the Weyl tensor, a so called Debever-Penrose (DP) direction. One of the main triumphs of
this effort, spread
out\footnote{see for example the reviews in \cite{GrifPod, Kramer}} between 1960 and 1980, has been the complete integration of the field equations (with a possible nonzero cosmological constant $\Lambda $), for the Petrov type D doubly aligned non-null EM fields, in which \emph{both} real PNDs
of $\mathbf {F}$ are parallel to a corresponding double DP vector and are geodesic as well as shear-free, the so called\cite{DebeverMcLen1981}
class $\mathcal{D}$ metrics\footnote{this class contains famous
examples such as the Reissner-Nordstr{\"o}m and Kerr-Newman solutions and, together with the Pleba\'{n}ski-Hacyan space-times\cite{PlebHacyan79}
and Garc{\'\i}a-Pleba\'{n}ski space-times\cite{GarciaPleban}, represents the
general solution for the doubly aligned Petrov type D EM fields}. In a recent study\cite{NVdB2017}
of non-aligned algebraically special EM fields it was noted that, at least for nonzero cosmological constant $\Lambda $, the double alignment condition of the class $\mathcal{D}$ metrics is actually a consequence of their multiple DP vectors being
geodesic and shear-free. Therefore this is also a necessary
condition for the existence of a 2-index Killing spinor, with the consequence of enabling\cite{WalkerPenrose70} to completely integrate the null geodesic equation for the
whole class $\mathcal{D}$. A natural question therefore arises as to whether EM solutions exist which are of Petrov type D, have $\Lambda  =0$ and in which the two real DP vectors $\mathbf {k} ,\mathbf {l }$ are geodesic and shear-free, but are
\emph{both non-aligned}\footnote{a related question for Petrov type
III was dealt with recently in \cite{NVdB2018}} with the PND's of a non-null electromagnetic field $\mathbf {F}$.
While the ``Kundt'' case of vanishing divergence of either $\mathbf {k}$ or $\mathbf {l }$ (i.e.~$\rho$ or $\mu=0$) can be dismissed, as it immediately implies at least half-alignment\footnote{one can also prove that ''half-Kundt'' 
necessarily implies ''double Kundt'' and hence double alignment}, the general case with $\rho \mu \neq 0$ remained elusive, even under
the simplifying ``double Robinson-Trautman'' (RT) assumption that $\mathbf {k}$ and $\mathbf {l }$ are both non-twisting. 

In this paper we give an affirmative answer to the above question. We present all corresponding double RT space-times satisfying the extra condition that the complex null vectors of the Weyl canonical tetrad are hypersurface orthogonal and discuss some of their properties.

The structure of the paper is as follows: in \S\ref{Main_eqs} we set up a suitable null tetrad, present the relevant Geroch-Held-Penrose\cite{GHP} (GHP) equations and show that the ``normalised'' Maxwell components $\frac{\zero}{\rho \pic}$ and $\frac{\two}{\mu \pi}$ are opposite complex numbers, allowing us to write
$\zero = \rho \pic \Czero \f$, $\two = - \mu \pi \Czero \f$ with $\f$ and $\Czero$ $(0,0)$-weighted GHP variables.\footnote{$\f$ positive and $| \Czero | = 1$} A completely integrable 
system is then constructed for the GHP variables describing the situation at hand. In \S\ref{MainRTeqs} we translate this into the corresponding Newman-Penrose (NP) variables. We then obtain a final system of partial differential equations and construct 
its general solution. 
In \S\ref{Discussion} some
properties of the resulting metrics are discussed.\\
Throughout we assume that the reader is familiar with the GHP and NP formalisms, but for convenience a short overview of GHP is presented in the Appendix.\\
For notations and sign conventions we refer to \cite{Kramer}.

\section{Main equations}\label{Main_eqs}
Investigating non-aligned Einstein-Maxwell fields first requires choosing an appropriate null tetrad, either adapting it to the Weyl tensor or to the electromagnetic field. Both approaches can have their advantages, but here, as we aim
to study non-null Einstein-Maxwell fields of Petrov type D, with an additional assumption on the DP vectors (namely their being geodesic and shear-free), it appears preferable to use a canonical Weyl tetrad.
The relevant equations are then obtained by substituting $\Psi_0=\Psi_1=\Psi_3=\Psi_4=0$, together with $\kappa=\nu=\sigma=\lambda=\rhoc  -\rho  =
\muc  -\mu = 0$ ($\mathbf {k} ,\mathbf {l }$ are assumed to be 
geodesic, shear-free and non-twisting) into equations (\ref{ghp1}-\ref{bi4}) of the Appendix. Note that we also impose the assumption $\tau+\pic=0$, which guarantees that the complex null vectors $\w{m}$ and $\overline{\w{m}}$ of the Weyl canonical tetrad are hypersurface orthogonal
($\w{m}\wedge \ud \w{m}=0$).\footnote{some preliminary work shows that large classes of solutions may exist when $\w{m}\wedge \ud \w{m}\neq 0$}
Next we define extension variables $\Eu= \tho \two,\Ev=\et \two,\EJ=-\tho'\rho$ ($\Eu, \Ev$ complex and $\EJ$ real), after which the Ricci, Bianchi and Maxwell equations (\ref{ghp1}-\ref{bi4}) are solved\footnote{to solve the Bianchi identities for the variables $\tho'\two,\etd \two$ it is essential that the electromagnetic field is non-null: $\zero \two -\one^2 \neq 0$} to yield the following system:

\begin{eqnarray}
\et\rho &= \zero \onec , \ 
\tho\rho = \rho^2+\zero \zeroc, \ 
\tho'\rho = -\EJ, \label{srho}\\
\et \pi &= -\pi \pic -\rho \mu +\EJ -\Psi_2 , \ 
\etd \pi = -\two \zeroc -\pi^2 , \ 
\tho' \pi = -\two \onec , \label{spi} \\
\et \one &= \mu \zero -2 \pic \one - \Eu', \ 
\tho'\one = -2 \mu \one +\pic \two + \Ev, \label{sone} \\
\et \two &= \Ev , \etd \two = 0, \ \tho \two = \Eu,\ \tho'\two = 0, \label{stwo}\\
\et \Psi_2 &= 2 \rho \one \twoc +2 \pic \one \onec -\twoc \Ev'+\onec \Eu'-3 \pic \psi ,\nonumber \\
\tho \Psi_2 &= 2 \rho \one \onec +2 \pic \one \zeroc -\onec \Ev'+\zeroc \Eu'+3 \rho \Psi_2,\ \nonumber \\
\tho' \Psi_2 &= -2 \mu \one \onec +2\pi \one \twoc + \onec \Ev -\twoc \Eu -3 \mu \Psi_2, \label{spsi}
\end{eqnarray}
with, by $(\ref{ghp5d})'-\overline{(\ref{ghp5d})}$, $\EJ-\EJ' = \overline{\Psi_2}-\Psi_2$ and hence ($\EJ, \EJ'$ being real) $\EJ-\EJ'=0$ and $\overline{\Psi_2}=\Psi_2$.
The equations for $\mu$ and $\zero$ have been omitted, as they can be obtained by ``priming'' equations (\ref{srho}) and (\ref{stwo}). Similarly $\etd \rho$, $\tho \pi$, $\etd \one$, $\tho \one$, $\etd \Psi_2$ and $\tho \Psi_2$ can be obtained by the prime and 
complex conjugation of (\ref{srho}-\ref{spsi}). This will hold throughout this section and results in a significant reduction of the computational effort.\\
Note that if $\rho, \mu$ or $\pi=0$, then by (\ref{srho},\ref{spi}) it immediately implies half-alignment\footnote{one can show that also double-alignment follows, i.e.~half-aligned Petrov type D Einstein-Maxwell-Kundt solutions with geodesic and shear-free DP vectors do not exist}.\\
We now apply the $\left[\etd,\et \right], \left[\et,\tho'\right],\left[\etd,\tho \right]$ commutators to $\two$ and the $\left[\etd,\tho'\right]$ commutator to $\one$ to obtain the following derivatives of $\Eu$ and $\Ev$,
\begin{eqnarray}
 \etd \Ev &= 2(\psi - \one \onec - \rho \mu)  \two,\label{d2v} \\
 \tho' \Ev &= 2(  \mu  \pic  -  \one  \twoc   )  \two-\mu \Ev, \label{d3v} \\
 \etd \Eu &= -2( \one \zeroc + \pi \rho )  \two-\pi \Eu,\label{d2u} \\
 \tho' \Eu &= ( 2 \pi \pic -2  \one \onec  + \Psi_2 )  \two-3 (\mu \Eu- \pi \Ev), \label{d3u} 
\end{eqnarray}
after which the $\left[ \tho', \tho\right]$ commutator applied to $\two$ results in an algebraic relation
\be
\two \Psi_2-\mu \Eu +\pi \Ev = 0 .\label{eq3}
\ee
We will use (\ref{eq3}) to express $\Eu$ and $\Ev$ in terms of the $(0,0)$-weighted quantity $\Ew = \Ev / \mu$. Evaluating the combination of commutators, $[\etd,\et]\one-[\etd,\tho'] \zero+[\et,\tho] \two$, we obtain the relation 
\begin{eqnarray}
\fl \pi \et  \Ew+ \pic \etd \EW  &= ( \EW+\Ew )(2 \rho \mu-\pi \pic -\EJ)+2 \zero \pi \mu +2 \two \pic \rho\nonumber \\
\fl &+\frac{\zero}{\rho} (2 \one-\Ew)( \zeroc   \mu-  \onec \pi)-\frac{\two}{\mu} ( 2 \one+\EW )(  \twoc   \rho+ \onec\pic)   \nonumber \\
\fl & +\psi (\frac {\zero \pi}{\rho} +{\frac {\two \pic}{\mu}} +\frac { \one \zero \zeroc}{\rho^2}
  -\frac {\one \two  \twoc }{\mu^2}), \label{eq5}
\end{eqnarray}
 which can be used to simplify the expression 
 $$
 \rho [\et,\tho'] \Ev' +\tau [\etd,\et] \Ev' +\Ev' [\etd, \et]\tau -\frac{\zero \psi +\tau \Ev'}{\rho} [\etd,\et]\rho  $$ to yield
 \begin{eqnarray}
 \fl \mu  \zeroc \et  \Ew -\rho  \twoc \etd \EW 
   &=  ( 2 \rho \mu+\psi )  ( \rho \mu-\psi )(\frac {\zero \zeroc }{\rho^2}- \frac{\two  \twoc }{\mu^2})+4 \one (\pi \rho \twoc+\pic\mu \zeroc) \nonumber \\
    & +  ( \EW+\Ew )( \onec \mu \rho +\onec \pi\pic+\one \zeroc \twoc)+2\one (\mu\rho-\pi\pic)(\Ewc+\EWc) \nonumber \\
  & +\pi\rho\twoc (3\EW+\Ew)-\pic \mu \zeroc (3 \Ew+\EW) \nonumber \\
  & +\pi\pic (\Ew\Ewc-\EW\EWc) +\rho \mu(\EW\Ewc-\Ew \EWc). \label{eq8}
 \end{eqnarray}
One can show that, when (\ref{eq5},\ref{eq8}), considered as a system for $\et \Ew$ and $\etd \EW$, is non-singular, solutions are necessarily doubly-aligned or conformally flat. We omit the tedious and lengthy proof of 
this property: the proof (in which the earlier derived reality of $\psi$ plays an essential role) is available from the authors, either by email to the second author for a semi-automated version, using the algebraic computing package STEM, or, via a more manual approach, from \cite{proof_Norbert}. When this system is singular, i.e.~when  
\be\mu \pi \zero + \rho \pic \two=0, \label{spec}
\ee
we can write 
 \be
\zero = \rho \pic \Czero \f,\  \two = - \mu \pi \Czero \f , \label{phi0and2}
\ee
with $(0,0)$-weighted
quantities $\f$ and $\Czero$, such that $\f$ is real positive and $|\Czero|=1$. 
Acting now on (\ref{spec}) with the operators $\pi \eth+ \pic \etd$ and $\pi \eth + \rho \thd$  results in
\be
\Ew+\EW = 0 \label{Wwrelation}
\ee
and
\be
(\rho \mu+\pi\pic)(\pi \zero \onec -\pic \one \zeroc)=0.
\ee
Rejecting the case $\rho \mu+\pi\pic=0$ (acting on this with the $\eth$ and $\thd$ operators immediately leads to conformal flatness), we have 
\be
\pi \zero \onec -\pic \one \zeroc = 0.\label{sbs}
\ee 
This allows us to define a (real positive) $(0,0)$-weighted function $\g$ by
\be
\one = \Czero \g, \label{sbsg}
\ee
which combining with the $\etd$ derivative of (\ref{spec}) and (\ref{sbs}) yields
\be
\EJ = \psi+\rho \mu (1+\f^2 \pi \pic)-\Czero \frac{\Ewc}{\f}. \label{Jexpr}
\ee
Finally, acting on (\ref{Wwrelation}) or (\ref{sbs}) leads to
\be
\et \Ew = 2 \Czero  \f \pic (\g^2 +\rho \mu -2\psi) - \Ew \pic \f \g .
\ee
We also note that (\ref{Jexpr}) shows that $\Czero \Ewc$ is real, so that we can write  $\Ew= \Czero \wzero$ with $\wzero$ real and $(0,0)$-weighted.
Application of $\et, \etd,\tho,\tho'$ to the $(0,0)$-weighted quantity $\Czero$ all return 0 and hence $\Czero$ is a constant. The only remaining variables are then $\f,\g,\wzero,\psi,\rho$ and $\pi$ (with $\f'=\f,\g'=\g,\wzero'=\wzero,\psi'=\psi$,
$\rho'=-\mu$, $\pi'=\pic$), which satisfy the completely integrable system
\begin{eqnarray}
\et \f &=-\pic \f (\f^2 \mu \rho+\f \g-1), \ \tho \f =-\f \rho (\f^2 \pi\pic -\f \g+1),\nonumber\\
\et \g &=\pic (\f \rho \mu-\f \psi-2 \g+\wzero), \ \tho \g =\rho (\f \pi \pic +2 \g-\wzero),\nonumber \\
\et \pi &= \pi\pic(\rho \mu \f^2 -1) -\frac{\wzero}{\f},\ \tho \pi = -\rho\pi \f \g ,\nonumber \\
\et \rho &= \rho \pic \f \g ,\ \etd \rho = \rho \pi \f \g,\nonumber\\
\tho \rho &= \rho^2(1+\f^2\pi\pic), \ \tho' \rho = -\rho\mu (\f^2\pi\pic +1) + \frac{\wzero}{\f} -\psi,\nonumber \\
\et \wzero &= \pic \f (2 \g^2-\g \wzero+2 \rho \mu-2 \psi),\ \tho \wzero = \f \pi \pic \rho (2 \f \g-\f \wzero+2),\nonumber \\
\et \psi &=-\pi (2\g \f \rho \mu-\f \rho \mu \wzero-\f \g \psi-2 \g^2+\g \wzero+3 \psi),\nonumber \\
\tho \psi &= \rho (\f^2 \pi\pic \psi+2\f \g \pi \pic -\f \pi \pic \wzero+2 \g^2-\g \wzero+3 \psi). \label{sysfinal}
\end{eqnarray}

\section{General solution of the double RT-case}\label{MainRTeqs}
We now use the previous results to set up an NP null tetrad $(\w{e}_a)=(\w{m},\overline{\w{m}}, \w{l},\w{k})$ with dual basis $\w{\omega}^a)$, construct an appropriate coordinate system and solve the field equations.
All results from the previous sections can be translated to the NP formalism
by means of the relations (\ref{GHP_NP}). In particular all relations involving only derivatives of the $(0,0)$-weighted GHP quantities carry over without modification.
In order to fix the null tetrad we use the fact that $\rho$ and $\mu$ are real and $\tau+\pic=0$, allowing one to specify a boost and spatial rotation such that $\pi$ and $\tau$ are real as well and
\begin{eqnarray}
 \mu &= e \rho, \quad (e=\pm 1),\\
 \pi &= -\tau.
\end{eqnarray}

$D \pi, \Delta \pi$ being real implies $\epsilon$ and $\gamma$ are real, while $\delta (\frac{\mu}{\rho})=0$ implies $\beta+\overline{\alpha}=0$.
From $D (\frac{\mu}{\rho})=\Delta (\frac{\mu}{\rho})=0$ and $\overline{\delta \pi}=\overline{\delta} \pi$ it follows then that the spin coefficients
$\alpha,\beta,\epsilon,\gamma$ are given by

\begin{eqnarray}
 \alpha &= -\beta = \wzero /(4 \pi \f),\\
 \gamma &= e \epsilon = (\f \psi - \wzero)/(4 \rho \f ).
\end{eqnarray}
Consequently the Cartan equations become
\begin{eqnarray}
 \ud \w{\omega}^1 &= \w{\omega}^1 \wedge ( -e \rho  \w{\omega}^3 + \rho \w{\omega}^4 + \frac{\wzero}{2\pi\f} \w{\omega}^2),  \label{Cartan1} \\
 \ud \w{\omega}^2 &= \w{\omega}^2 \wedge ( -e \rho  \w{\omega}^3 + \rho \w{\omega}^4 + \frac{\wzero}{2\pi\f} \w{\omega}^1), \\
 \ud \w{\omega}^3 &= \w{\omega}^3 \wedge ( -\pi \w{\omega}^1 -\pi \w{\omega}^2 + e \frac{\wzero-\f \psi }{2\rho \f}\w{\omega}^4),\\
 \ud \w{\omega}^4 &= \w{\omega}^4 \wedge ( -\pi \w{\omega}^1 -\pi \w{\omega}^2 - \frac{\wzero-\f \psi }{2\rho \f}\w{\omega}^3), \label{Cartan4}
\end{eqnarray}
showing that the basis vectors are all hypersurface-orthogonal. From (\ref{Cartan1}-\ref{Cartan4}) it is clear that this also holds for the basis dual to the one-forms $e \w{\omega}^3+\w{\omega}^4, \w{\omega}^1-\w{\omega}^2,
\nO^1=e \w{\omega}^3-\w{\omega}^4, \nO^2=\w{\omega}^1+\w{\omega}^2$, the latter two of which satisfy
\be
 \ud \nO^1 = - \pi \nO^1 \wedge \nO^2, \  \ud \nO^2 = \rho \nO^1 \wedge \nO^2. \label{O1O2eq}
\ee
Next we introduce new variables $\h = \wzero-2\g$ and $\j = \f\psi + 2\g - \wzero$, which 
simplify\footnote{using, for example, $\ud \f = \w{\omega}^1 \delta \f+\w{\omega}^2 \overline{\delta} \f +\w{\omega}^3 \Delta \f + \w{\omega}^4 D \f$ etc.} 
the system (\ref{sysfinal}) to
\begin{eqnarray}
 \ud \f &= \f [(\f^2\pi^2\rho - \f\g\rho + \rho)\nO^1 + (-e \f^2\pi\rho^2 - \f\g\pi + \pi) \nO^2],\label{n_eqf}\\
 \ud \g &= -\nO^1 (\f \pi^2 - \h)  \rho + \pi (e \f  \rho^2 - \j) \nO^2, \label{n_eqg}\\
 \ud \h &= \h [ (\f^2 \pi^2  \rho - 2  \rho) \nO^1 - (\f \g + 2) \pi \nO^2], \label{n_eqh}\\
 \ud \j &= \j[ - (\f \g + 2)  \rho \nO^1 + (-e \f^2 \pi  \rho^2 - 2 \pi) \nO^2], \label{n_eqj}\\
 \ud  \rho &= -\frac{1}{2 e \f}(2 e \f^3 \pi^2  \rho^2 + 2 e \f  \rho^2 - 2 \g + \j) \nO^1 + \nO^2 \f \g \pi  \rho, \label{n_eqrho}\\
 \ud \pi &= \frac{1}{2 \f} (2 e \f^3 \pi^2  \rho^2 - 2 \f \pi^2 - 2 \g - \h) \nO^2 + \nO^1 \f \g \pi  \rho. \label{n_eqpi}
\end{eqnarray}
The null tetrad being fixed and $\f,\g,\h,\j,\rho,\pi$ being the remaining (non-constant) spin coefficients and (suitably transformed) Maxwell and curvature components, (\ref{n_eqf}-\ref{n_eqpi}) show that this set contains at most two functionally independent functions and hence the corresponding space-times
will admit at least two Killing vectors. One can actually show that this is the maximally allowed number, as the vanishing of all double wedge products would lead to an inconsistency (this will
be obvious from the explicit solutions as well).

Introducing coordinates $\x,\y,\u,\v$ such that
\begin{eqnarray}
 & \w{\omega}^1-\w{\omega}^2 = i \nP \ud \y,\ e \w{\omega}^3+\w{\omega}^4 =\nQ \ud \x,\label{PQdef} \\
  & \nO^1 = \nB \ud \u, \  \nO^2 = \nC \ud \v, \label{BCdef}
\end{eqnarray}
($\x$ and $\y$ clearly then being the ignorable coordinates corresponding to the two Killing vectors) it follows from (\ref{O1O2eq}) that
\be
\nB_{,\u} =  \pi \nB \nC, \ \nC_{,\v} = \rho \nB \nC, \label{rhopisol}
\ee
after which a linear combination of (\ref{n_eqrho})and (\ref{n_eqpi}) shows that $(\log(\nB / \nC))_{,\u \v}=0$. Hence $\nB / \nC$ is separable in $\u$ and $\v$ and a coordinate transformation exists such that
(re-defining $\nB$ and $\nC$) $\nB = \nC$. The exterior derivatives of (\ref{PQdef}) lead then to two partial differential equations for $\nP$ and $\nQ$,

\begin{eqnarray}
 \ud \log (\nP {\nP}_0) &= -\frac{\nB}{2\pi\f} (\h + 2\g)\ud \u  + \nB \rho \ud \v, \label{PQdefbis1} \\
 \ud \log (\nQ {\nQ}_0) &= \frac{e \nB}{2\rho\f} (2\g - \j)\ud \v + \nB \pi \ud \u , \label{PQdefbis2}
\end{eqnarray}
with ${\nP}_0= {\nP}_0 (\y)$ and ${\nQ}_0 = {\nQ}_0 (\x)$.
Without loss of generality 
one can put ${\nP}_0 = {\nQ}_0 =1$, such that, by (\ref{n_eqrho},\ref{n_eqpi}), the previous relations can be rewritten as

\begin{eqnarray}
\ud \log (\frac{\nQ }{\nB \rho} ) &= \nB \f \pi (\rho \pi \f \ud \v - \g \ud \u), \label{PQdefbis3} \\
\ud \log (\frac{\nP}{\nB \pi} )  &= -\nB  \f \rho (e \rho \pi \f \ud \u  + \g \ud \v). \label{PQdefbis4}
\end{eqnarray}

At this point we will make a distinction between the cases $\h \j \neq 0$ (i.e. 
$(\delta \Phi_2 - \mu \Phi_1)( \delta \Phi_2 - \mu \Phi_1 -\frac{\mu}{2 \rho \pi}\Psi_2 \Phi_0) \neq 0$) and $\h=0$ ($\delta \Phi_2 - \mu \Phi_1=0$) or $\j=0$ ($\delta \Phi_2 - \mu \Phi_1 -\frac{\mu}{2 \rho \pi}\Psi_2 \Phi_0 = 0$). Only the first two will be treated in detail below, 
as the analysis of the case $\j=0$ is essentially identical to that of $\h=0$, since the transformation 
\begin{eqnarray}
(\f,\g,\h,\j,\pi,\rho) &\rightarrow (\f, \g, -\j, - \h,  i \rho / \sqrt{e} , i \pi \sqrt{e}) , \nonumber \\
(\nO^1,\nO^2) &\rightarrow (  i \nO^2 / \sqrt{e}, -i \nO^1 \sqrt{e}) , \label{invariance}
\end{eqnarray}
leaves the
system (\ref{n_eqf}-\ref{n_eqpi}) invariant. Note that $\h=\j=0$ is excluded, as it would imply
either conformal flatness ($\Psi_2=0$) or double alignment ($\f=0$).
\subsection{The case $\h \j \neq0$}\label{hjnonzero}
When $\h \j \neq 0$ one can use (\ref{n_eqh},\ref{n_eqj}) for integrating (\ref{n_eqf}) and (\ref{PQdefbis3},\ref{PQdefbis4}) to obtain
\be
\nP = \j \pi \nB^3,\ \nQ = \h \rho \nB^3\label{PQdef3}
\ee
and
\be
\f = \f_0 \h \j \nB^5.\label{fzerodef}
\ee
Here $\f_0$ is an integration constant, which we will put $= 1$ by a global re-scaling of the metric\footnote{we note that the system (\ref{n_eqf}-\ref{n_eqpi}) is invariant under the transformation $\ud s^2 \to a^2 \ud s^2$ (and hence
$\w{\omega}^b \to a\, \w{\omega}^b$ for the dual basis vectors of the NP null tetrad), as this implies $\f \to a \f$, $\nB \to a \nB$ and
$(\pi,\rho,\j,\h,\g) \to a^{-1} (\pi,\rho,\j,\h,\g)$}.\\
Now (\ref{n_eqh},\ref{n_eqj}) imply
\begin{eqnarray}
 \ud \log (\h \rho  \nB^2) &= \frac{2 \g -\j -2 e \h\j \nB^5 \rho^2}{2 e \h\j \nB^4 \rho} \ud \v ,\\
 \ud \log (\j \pi \nB^2) &= -\frac{2 \g +\h +2 \h\j \nB^5 \pi^2}{2  \h\j \nB^4 \pi}\ud \u ,
\end{eqnarray}
showing that functions $\nvarsigma=\nvarsigma(\u), \nxi=\nxi(\v)$ exist such that
\be
\pi = \frac{\nvarsigma}{\j \nB^2}, \ \rho = \frac{\nxi}{\h \nB^2} \label{rhopiexpr2}
\ee
and
\begin{eqnarray}
\g &= e \j \nB^2 \nxi' + e \frac{\j \nB \nxi^2}{\h} + \half \j \label{gsola}  \\
   &=  - \h \nB^2 \nvarsigma' - \frac{\h \nB \nvarsigma^2}{\j} - \half \h, \label{gsolb}
\end{eqnarray}
where we have written $\nxi', \nvarsigma',\xi'', \ldots $ for the derivatives of $\nxi$ and $\nvarsigma$ w.r.t.~$\u$ and
$\v$.
Subtracting (\ref{gsolb}) from (\ref{gsola}) reveals the following key algebraic relation between $\j$ and $\h$,
\be
\nB (\h^2\nvarsigma^2 + e \j^2\nxi^2) + (\nB^2\nvarsigma' + \half)\j\h^2 + (e \nB^2 \nxi' + \half)\h\j^2  = 0, \label{key1}
\ee
while the expressions for (\ref{PQdef3}) reduce to
\be
\nP = \nB\nvarsigma,\ \nQ=\nB\nxi . \label{PQdef4}
\ee
The metric then becomes
\be
\ud s^2 = \frac{\nB^2}{2} ( \ud \u^2 + e \ud \v^2 - e \nxi^2 \ud \x^2   + \nvarsigma^2\ud \y^2 ) . \label{ds2_1}
\ee
With the introduction of new variables $\nN,\nJ,\nH$ by
$$\j=\nJ/ \nB,\  \h=\nH / \nB,\  \nB=\nN^{-1/2},$$ equations ({\ref{key1}) and (\ref{n_eqh},\ref{n_eqj}) simplify to
\be
 \nH  \nJ ( \nH +  \nJ)  \nN + 2  e  \nH  \nJ^2  \nxi' + 2  e  \nJ^2  \nxi^2 + 2  \nH^2  \nJ  \nvarsigma' + 2  \nH^2  \nvarsigma^2 = 0 , \label{KEY1}
\ee
\begin{eqnarray}
\fl  \ud \nJ &= - \frac{\nJ}{2  \nN^2  \nH}  \nxi (2  e  \nH  \nJ^2  \nxi' + 2  e  \nJ^2  \nxi^2 +  \nH  \nJ^2  \nN + 2  \nN^2) \ud \v -  \frac{\nvarsigma}{\nN^2 } ( e  \nJ^2  \nxi^2 +  \nN^2) \ud \u ,\label{EQ1}\\
\fl  \ud \nH &= - \frac{\nvarsigma}{2  \nN^2  \nJ}  \nH (2  e  \nH  \nJ^2  \nxi' + 2  e  \nJ^2  \nxi^2 +  \nH  \nJ^2  \nN + 2  \nN^2) \ud \u +  \frac{\nxi}{\nN^2} ( \nH^2  \nvarsigma^2 -  \nN^2) \ud \v  ,\label{EQ2}
\end{eqnarray}
whereas (\ref{rhopisol}) implies
\be
\ud \nN = -2  \nN (\frac{\nxi}{\nH} \ud \v   + \frac{\nvarsigma}{\nJ} \ud \u ). \label{EQ0}
\ee
A second algebraic equation is now obtained from (\ref{n_eqg}) and (\ref{gsola}),
\begin{eqnarray}
\fl & (4  e  \nH^2  \nJ^2  \nxi'' + 16  e  \nH  \nJ^2  \nxi  \nxi' + 12  e  \nJ^2  \nxi^3 -  \nH^2  \nJ^4  \nxi + 8  \nH^2  \nJ  \nxi  \nvarsigma' + 12  \nH^2  \nxi  \nvarsigma^2)  \nN^2 \nonumber \\
\fl & -  4  \nH  \nJ^4  e  \nxi ( \nH  \nxi' +  \nxi^2)  \nN - 4  \nJ^2  \nxi ( e  \nH^2  \nxi^2  \nvarsigma^2 +  \nH^2  \nJ^2  {\nxi'}^2
 + 2  \nH  \nJ^2  \nxi^2  \nxi' +  \nJ^2  \nxi^4)=0 \label{KEY2}
\end{eqnarray}
or, using (\ref{gsolb}) instead of (\ref{gsola}),
\begin{eqnarray}
\fl & ( 4  \nH^2  \nJ^2  \nvarsigma'' + 16  \nH^2  \nJ  \nvarsigma  \nvarsigma' + 12  \nH^2  \nvarsigma^3)  \nN^2 +  \nH^4  \nJ^2  + 8  e  \nH  \nJ^2  \nxi'  \nvarsigma + 12  e  \nJ^2  \nxi^2  \nvarsigma  \nvarsigma  \nonumber \\
\fl & + 4  \nH^4  \nJ  \nvarsigma ( \nJ  \nvarsigma' +
 \nvarsigma^2)  \nN + 4  \nvarsigma  \nH^2 ( e  \nJ^2  \nxi^2  \nvarsigma^2 +  \nH^2  \nJ^2  {\nvarsigma'}^2 + 2  \nH^2  \nJ  \nvarsigma^2  \nvarsigma' +  \nH^2  \nvarsigma^4) = 0, \label{KEY3}
\end{eqnarray}
an equation which also can be obtained by taking the exterior derivative of (\ref{KEY1}).
Eliminating the first derivatives of $\nxi, \nvarsigma$ from (\ref{KEY2},\ref{KEY3}) yields
\be
  e  \frac{\nxi''}{\nxi} +  \frac{\nvarsigma''}{\nvarsigma} = 3  \nN ( \frac{1}{\nJ} + \frac{1}{\nH}). \label{KEY23}
\ee
Taking the exterior derivative of this equation leads to one more algebraic relation between $\nJ,\nH$ and $\nN$,
\be
 (\frac{1}{\nJ^2} - \frac{1}{\nH^2})  \nN^2 -\frac{1}{3} ( \nCHI  e +  \nSigma)  \nN -  e  \nxi^2 -  \nvarsigma^2 = 0, \label{KEY4}
\ee
where we have defined
\be
\nCHI= \frac{\nxi'''}{\nxi^2}- \frac{\nxi'  \nxi''}{ \nxi^3}, \textrm{ and }  \nSigma =   \frac{\nvarsigma'''}{ \nvarsigma^2}- \frac{\nvarsigma'  \nvarsigma''}{\nvarsigma^3}.\label{defCHISigma}
\ee
The exterior derivative of (\ref{KEY4}) now yields two ODE's,
\be
 \nCHI'  e - 3  \nxi = 0 =  \nSigma' + 3  \nvarsigma, \label{CHISigmaODE}
\ee
first integrals of which are given by
\be
\varsigma'' = -\frac{\nvarsigma}{6}\nSigma^2 + 3 \nSigma_0 \nvarsigma \textrm{ and } \nxi'' = e\frac{\nxi}{6}\nCHI^2 + 3 \nCHI_0 \nxi \label{firstints}
\ee
($\nSigma_0,\nCHI_0$ constants).
Taking succesive derivatives of the components of (\ref{EQ0}) and using (\ref{EQ1},\ref{EQ2}), we obtain two linear equations for $\nNu$ and $\nNv$,
\begin{eqnarray}
 \nN_{,\u\u\u} &= -\nNu (\frac{1}{6}\nSigma^2 -3 \nSigma_0 + 3\frac{{\nvarsigma'}^2}{\nvarsigma^2}) + 3\nNuu\frac{\nvarsigma'}{\nvarsigma},\\
 \nN_{,\v\v\v} &=  \nNv(\frac{e}{6} \nCHI^2 + 3  \nCHI_0 - 3 \frac{{\nxi'}^2}{\nxi^2}) + 3\nNvv\frac{\xi'}{\nxi},
\end{eqnarray}
the general solutions of which are given by
\begin{eqnarray}
 \nNu &= F_1(\v) \nvarsigma + F_2(\v) \ns \, \nvarsigma,\label{Nu}\\
 \nNv &= F_3(\u)\nxi + F_4(\u)\nx \, \nxi,\label{Nv}
\end{eqnarray}
with $F_1, \ldots F_4$ being arbitrary functions of $\u$ or $\v$ and where we have defined $\ns$ and $\nx$ by 
\be
\nvarsigma = \ns' , \ \nxi= \nx' . \label{xi0sigma0def}
\ee
The integrability conditions for (\ref{Nu},\ref{Nv}) then show that $F_1, \ldots F_4$ must be quadratic functions of $\u$ or $\v$. Herewith (\ref{Nu},\ref{Nv}) can be integrated to yield
\be
\nN = c_1 \nx^2\ns^2 + c_2\nx \ns^2 + c_3 \nx^2 \ns + c_4 \ns^2 + c_5 \nx \ns + c_6 \nx^2 + c_7 \ns + c_8 \nx + c_9 .\label{N_expression}
\ee
Substituting this into (\ref{EQ0}) gives expressions for $\nJ$ and $\nH$,
\begin{eqnarray}
\nJ &= -2 \nN (2  c_1  \nx^2  \ns + 2  c_2  \nx  \ns +  c_3  \nx^2 + 2  c_4  \ns +  c_5  \nx +  c_7)^{-1}, \\
\nH &= -2 \nN (2  c_1  \nx  \ns^2 +  c_2  \ns^2 + 2  c_3  \nx  \ns +  c_5  \ns + 2  c_6  \nx +  c_8)^{-1},
\end{eqnarray}
which, together with (\ref{KEY23}), imply $c_1=0, c_2=1, c_3=-1$, and
\be
 \nCHI = 3 e (\nx  - c_6 - \frac{c_5}{2}), \textrm{ and } \nSigma = - 3 (\ns + c_4 + \frac{c_5}{2}), \label{CHISIGMAexpr}
\ee
together with
\begin{eqnarray}
  & [ 3   \nx^2 - 3 (c_5  + 2 c_6  ) \nx + c_{10} - 3   c_7] \nxi - 2  e   \nxi'' = 0, \label{xicond1}\\
 & [ 3   \ns^2 +3  ( c_5  + 2  c_4  ) \ns  +  c_{10} + 3   c_8]   \nvarsigma + 2   \nvarsigma'' = 0 \label{varsigmacond1}.
\end{eqnarray}

Combining (\ref{CHISIGMAexpr},\ref{xicond1},\ref{varsigmacond1}) with (\ref{EQ1},\ref{EQ2}) leads to
\be
c_{10} = -2 c_4 c_6 + \half c_5^2 + 2c_7 - 2 c_8,
\ee
and two quadratures determining $\nvarsigma, \nxi$ as functions of $\u$ and $\v$:
\begin{eqnarray}
 4 e \nxi^2 &=  \nx^4 - 2 (c_5 + 2  c_6) \nx^3 - (4  c_4  c_6 - c_5^2 + 2  c_7 + 4  c_8) \nx^2 \nonumber \\
 & - 2 (2  c_4  c_8 - c_5  c_7 + 2  c_9) \nx - 4  c_4  c_9 + c_7^2, \label{xisq1} \\
 4 \nvarsigma^2 &= -\ns^4 - 2 (c_5+2 c_4 )  \ns^3 + (4 c_4  c_6 - c_5^2 - 2 c_8- 4 c_7 )  \ns^2 \nonumber \\
 & + 2 (2 c_6  c_7 - c_8  c_5 - 2 c_9)  \ns + 4 c_6  c_9 - c_8^2. \label{varsigmasq1}
\end{eqnarray}
One can adjust the constants $c_4$ and $c_6$ by means of a translation of $\ns$ and $\nx$. Specifically we can choose $- 2 c_4=- 2 c_6= c_5\equiv\np$ and $c_9\equiv\nq$, so 
that (\ref{CHISIGMAexpr}) reduces to
\be
\nCHI= 3  e \nx, \textrm{ and } \nSigma = - 3 \ns . \label{CHISIGMAexpr2}
\ee
Replacing $c_7,c_8$ by 
\be
3 \nSigma_0 = -\frac{c_8}{2} - c_7,\  3 e \nCHI_0 =   -\frac{c_7}{2} + c_8,
\ee
the relations (\ref{N_expression},\ref{xisq1},\ref{varsigmasq1}) simplify to
\begin{eqnarray}
 \nN &= \nx \ns^2-\ns\nx^2-\frac{\np}{2}(\nx - \ns)^2   - 2 e (2 \nx - \ns) \nCHI_0 + 2 ( \nx - 2 \ns) \nSigma_0  +  \nq ,\label{N_expressionfinal}\\
 \nxi^2 &= \smfrac{e}{4} \nx^4 + 3 \nCHI_0 \nx^2 - (\np(e \nSigma_0  + \nCHI_0) + e  \nq ) \nx + e (2 \nSigma_0 - \nCHI_0)^2 + \half e  \np  \nq , \label{xisimp}  \\
 \nvarsigma^2 &= -\smfrac{1}{4}\ns^4 + 3 \nSigma_0 \ns^2 + (\np(e \nCHI_0  + \nSigma_0) -  \nq) \ns  - (2\nCHI_0-e \nSigma_0)^2 - \half \np  \nq ,\label{sigsimp}
\end{eqnarray}
$\np,\nq,\nSigma_0,\nCHI_0$ being independent constants of integration, where $\nSigma_0,\nCHI_0$ are the two conserved quantities  introduced in (\ref{firstints}).
Using $\ns$ and $\nx$ as coordinates instead of $\u$ and $\v$ and re-introducing a global scale-factor $k^2$ (which we used in (\ref{fzerodef}) to put the integration constant $f_0=1$), the metric (\ref{ds2_1}) finally reads
\be
\ud s^2 = \frac{k^2}{2 \nN} ( \nvarsigma^{-2} {\ud \ns}^2 + e \nxi^{-2} {\ud \nx}^2 - e \nxi^2 \ud \x^2   + \nvarsigma^2\ud \y^2 ), \label{ds2_2}
\ee
with $\nN$ given by (\ref{N_expressionfinal}) and $\nxi,\nvarsigma$ by (\ref{xisimp},\ref{sigsimp}).

\subsection{The case $\h=0$}\label{hzero}
When $\h=0$ the expressions for $\nP$ and $\pi$ in (\ref{PQdef3},\ref{rhopiexpr2},\ref{PQdef4}) and for $\g$ in (\ref{gsola}) are obtained as in the previous section, but (\ref{n_eqj},\ref{n_eqpi}) now
provide the following algebraic restriction on $\f,\j$ and $\nB$,
\be
(2\nB\j\nvarsigma' + 2\nvarsigma^2)\f^2 + \nB^4\j^3(2 e \nB^2 \nxi' + 1)\f + 2e \nB^{10} \j^4\nxi^2 = 0. \label{h0key1}
\ee
However, instead of (\ref{rhopiexpr2}b), we use the condition $\frac{1}{\j}\times$(\ref{n_eqj})-$\frac{1}{\f}\times$(\ref{n_eqf})-$\frac{1}{\rho}\times$(\ref{n_eqrho}), which now implies
\be
\rho = \frac{\nxi \j}{\f} \nB^3.
\ee
In this case (\ref{PQdefbis2}) reduces to $\ud \log \left[\nQ / (\nB \nxi) \right] =0$, so that again (\ref{PQdef4}) holds.\\
Two more algebraic relations between $\f,\j$ and $\nB$ are obtained by substituting (\ref{gsola}) in (\ref{n_eqg}) and into the exterior derivative of (\ref{n_eqf}), thereby leading to
\begin{eqnarray}
\fl  & [- \nB^4  \nxi (4  \nB^4   \nxi'^2 + 4  \nB^2 e   \nxi' + 1)  \j^4 - 4  \nB^2 e ( \nB^4  \nxi^3  \nvarsigma^2 -   \nxi'')  \j^2 + 4  \nvarsigma^2  \nxi]  \f^2 \nonumber \\
\fl &+ [-4  \nB^{10}  \nxi^3 (2  \nB^2   \nxi' + e)  \j^5  + 4  \nB^4  \nxi (2  \nB^2 e   \nxi' - 1)  \j^3]  \f
  - 4  \nB^{16}  \j^6  \nxi^5 + 4  \nB^{10} e  \j^4  \nxi^3 = 0, \label{h0key2} \\
\fl & ( \nB^2  \j^2  \nvarsigma   \nvarsigma'^2 + 2  \nB  \j  \nvarsigma^3   \nvarsigma' +  \nvarsigma^5)  \f^4 + [ \nB^6 ( \nB^4 e  \nxi^2  \nvarsigma^3 +   \nvarsigma'')  \j^4 + 2  \nB^5  \j^3  \nvarsigma   \nvarsigma'
 +  \nB^4  \j^2  \nvarsigma^3]  \f^2 \nonumber \\
\fl  & -  \nB^8  \f  \j^5  \nvarsigma +  \nB^{14} e  \j^6  \nxi^2  \nvarsigma = 0. \label{h0key3}
\end{eqnarray}
Again we introduce new variables $\nN, \nJ, \nF$ by $\nB=\nN^{-1/2}, \j=\nJ/\nB^3,\f=\nJ \nF/\nB$ and combine (\ref{h0key1},\ref{h0key2},\ref{h0key3}) to obtain
\begin{eqnarray}
 &  \nJ^2  \nF  \nN^2 + 2  \nF  \nJ (e  \nJ  \nxi' +  \nF   \nvarsigma')  \nN + 2 e  \nJ^2  \nxi^2 + 2  \nF^2  \nvarsigma^2 = 0, \label{h0KEY1}\\
&   \nJ^2  \nF \frac{ \nF^3  \nxi   \nvarsigma'^2 - e  \nF  \nxi'' + 2  \nxi}{ \nF^2  \nvarsigma^2 + 1}  \nN^2 + 2  \nJ   \nvarsigma'  \nF^2   \nxi  \nN 
  +   \nxi  ( e \nJ^2  \nxi^2 +  \nF^2  \nvarsigma^2) = 0,\label{h0KEY2}\\
& \frac{e  \nxi''  \nvarsigma +  \nxi   \nvarsigma''}{ \nxi  \nvarsigma} - \frac{3}{\nF} = 0,\label{h0KEY3}
\end{eqnarray}
while the partial differential equations for $\nB,\f,\j$ become
\begin{eqnarray}
\fl \ud \nN &=  -2 \frac{\nxi}{\nF} \ud \v  - 2 \frac{\nvarsigma}{\nJ} \ud \u   \label{h0dN}, \\
\fl \ud \nF &= -\frac{\nF}{2  \nJ  \nN}  \nvarsigma  (2  \nF  \nJ^2  \nN  e  \nxi' + \nF  \nJ^2  \nN^2 + 2  \nJ^2  e  \nxi^2 - 2)  \ud \u \nonumber \\
\fl  & + \frac{\nxi}{2  \nN}  (2  \nF  \nJ^2  \nN  e  \nxi' + 2  \nF^2  \nJ  \nN  \nvarsigma' + \nF  \nJ^2  \nN^2 + 2  \nJ^2  e  \nxi^2 + 4  \nF^2  \nvarsigma^2 + 2)  \ud \v \label{h0dF}\\
\fl \ud \nJ &= -\frac{\nvarsigma}{\nN}  (\nJ^2  e  \nxi^2 - 1)  \ud \u - \frac{\nxi  \nJ}{2  \nF  \nN}  (2\pic  \nF  \nJ^2  \nN  e  \nxi' + \nF  \nJ^2  \nN^2 + 2  \nJ^2  e  \nxi^2 - 2)  \ud \v \label{h0dJ} .
\end{eqnarray}
Herewith (and with the quantities $\nCHI,\nSigma$  defined by (\ref{defCHISigma})), the exterior derivatives of (\ref{h0KEY2},\ref{h0KEY3}) yield
\begin{eqnarray}
 \nN &= -\frac{3 e  ( \nF^2  \nvarsigma^2 + 1)}{ \nCHI  \nF^2}, \label{h0KEY2d} \\
 \nJ &= \frac{ e  \nF  \nCHI}{3  \nF   \nvarsigma' + \nSigma}. \label{h0KEY3d}
\end{eqnarray}
While the exterior derivative of (\ref{h0KEY2d}) becomes an identity under (\ref{h0KEY1}-\ref{h0KEY3d}), the exterior derivative of (\ref{h0KEY3d}) results in (compare with (\ref{CHISigmaODE}))
\be
\nSigma' + 3  \nvarsigma = 0 \textrm{ and } \nCHI'=0. \label{CHISigmaODEbis}
\ee
One therefore again obtains the first integral for $\nvarsigma$ as in (\ref{firstints}), but now this is complemented by $\nCHI=\nCHI_0$,  where $\nCHI_0$ is
an integration constant (with $\nCHI_0 \neq 0$ by (\ref{h0KEY2d}) ). \\
As in section \ref{hjnonzero} one simplifies (\ref{h0KEY2d}) with (\ref{h0KEY3}) and (\ref{h0KEY3d}), to obtain a partial differential equation for $\nN$, which
can be integrated to yield (with $\ns, \nx$ defined as in (\ref{xi0sigma0def}))
\begin{eqnarray}
12 e  \nCHI_0 \nN &= -(3 \ns^2 -2 e \nCHI_0 \nx)^2 + (-6 c_1 e + 36  \nSigma_0) \ns^2  - 4  \nCHI_0 (6 e \nSigma_0  - c_1)  \nx \nonumber\\
& - 36 \nvarsigma^2 + 12 e \nSigma_0 c_1  - 36  \nSigma_0^2 - c_1^2, \label{h0tempN}
\end{eqnarray}
together with the condition
\be
 -\nCHI_0 \nx^2 + c_1 \nx + c_0 + 2 \nxi' = 0. \label{h0_xi_equation}
\ee
Now (\ref{h0dF},\ref{h0dJ}) determine $\nF,\nJ$ and by substituting these into the equations (\ref{h0KEY1}-\ref{h0dJ}) one further obtains restrictions on the functions $\nxi$ and $\nvarsigma$. Using a translation of $\nx$ to put the
integration constant $c_1=0$, together with some tedious algebra, eventually leads to the following relations:
\begin{eqnarray}
\nxi^2 &= \frac{\nCHI_0}{3} \nx^3 - \npp \nx +\frac{3 e}{4\nCHI_0^2}(8\nCHI_0\nSigma_0 \npp - 96\nSigma_0^3 + 3\nK^2) \label{h0xi_equation_bis},\\
\nvarsigma^2 &= -\smfrac{1}{4}s^4 + 3\nSigma_0\ns^2 - \smfrac{\npp}{3}\nCHI_0 - \nK \ns + 3\nSigma_0^2 , \label{h0s_eq3bis}
\end{eqnarray}
where $\npp, \nK$ are new constants of integration.\\
Using $\ns , \nx$ as coordinates instead of $\u,\v$ the metric remains as given by (\ref{ds2_2}), but now
$\nxi$ is given by (\ref{h0xi_equation_bis}), while (\ref{h0tempN}) reduces to
\be
\nN = (\nx - 6 e\frac{\nSigma_0}{\nCHI_0})\ns^2 + 3 \frac{e\nK}{\nCHI_0} \ns - \frac{e \nCHI_0}{3} \nx^2 - 2\nSigma_0\nx + e \frac{\nCHI_0 \npp - 12\nSigma_0^2}{\nCHI_0} .\label{h0Nexp}
\ee

\subsection{The case $\j=0$}\label{jzero}
As the analysis of the case $\j=0$ is almost identical to that of $\h=0$ (cf.~the invariance  of the system (\ref{n_eqf}-\ref{n_eqpi}) under the transformation (\ref{invariance})), we limit ourselves to only presenting the
results.\\
The metric is still given by (\ref{ds2_2}), but now $\nxi$ and $\nvarsigma$ read
\begin{eqnarray}
 \nxi^2 &= \smfrac{e}{4} \nx^4  + 3\nCHI_0 \nx^2  + \smfrac{\npp}{3} \nSigma_0 + \nK\nx - 3 e \nCHI_0^2, \label{j0x_eq3bis} \\
 \nvarsigma^2 &= \frac{\Sigma_0}{3} \ns^3 - e \npp \ns + \frac{3}{4\nSigma_0^2}( 8 \nCHI_0\nSigma_0 \npp - 96 e \nCHI_0^3- 3\nK^2), \label{j0s_eq3bis}
\end{eqnarray}
while (\ref{h0Nexp}) is replaced by
\be
\nN = (6 e \frac{\nCHI_0}{\nSigma_0} - \ns) \nx^2 - \frac{\nSigma_0}{3} \ns^2 + 3 \frac{e\nK}{\nSigma_0} \nx  - 2 e \nCHI_0 \ns + \frac{e \nSigma_0 \npp - 12\nCHI_0^2}{\nSigma_0}. \label{j0Nexp}
\ee

\section{Discussion}\label{Discussion}
We have constructed all Petrov type D Einstein-Maxwell fields of Robinson-Trautman type (i.e.~with expanding but non-twisting DP vectors) in which the Maxwell field is totally non-aligned with the DP vectors and in which the latter are
assumed to be geodesic and shear-free, with $\w{m}$ being hypersurface orthogonal. All these solutions necessarily have a vanishing cosmological constant and are given by the metric (\ref{ds2_2}). Three different 5-parameter classes exist:
\begin{itemize}
\item $\h \j \neq 0$ with $\nN,\nvarsigma,\nxi$ given by (\ref{N_expressionfinal},\ref{sigsimp},\ref{xisimp}),
\item $\h=0, \j \neq 0$ with $\nN,\nvarsigma,\nxi$ given by (\ref{h0Nexp},\ref{h0s_eq3bis},\ref{h0xi_equation_bis}),
\item $\j=0, \h \neq 0$ with $\nN,\nvarsigma,\nxi$ given by (\ref{j0Nexp},\ref{j0s_eq3bis},\ref{j0x_eq3bis}),
\end{itemize}
In all cases the electromagnetic field is given by $$\Phi_0=-\Phi_2=\Czero\nxi\varsigma\nN^{-\half}k^{-1}, \ \Phi_1=\Czero\g k^{-1}$$ with $\g$ a not very illuminating expression obtainable from (\ref{gsola}) or (\ref{gsolb}).

We note that all solutions with $e=+1$ are static in the domain where $\nN, \nvarsigma,\nxi$ are positive, with time-like Killing vector $\partial_{\x}$. The Einstein-Maxwell equations for static space-times in which both electrostatic and magnetostatic
fields are present have been investigated in \cite{Das1979}, where it was proved that the electric and magnetic field vectors $\w{E}$ and $\w{B}$ (evaluated w.r.t. the time-like Killing vector) must be parallel. This is consistent with our results, as $E_a+ i B_a =\sqrt{2}[ \Phi_2 m_a
-\Phi_0 \overline{m}_a+\Phi_1 (k_a-l_a)]$, with $\Phi_0,\Phi_1,\Phi_2$ all having the same (constant) phase factor.

From the general expression
\footnote{note that this is less obvious when using the simplified form given by (\ref{N_expressionfinal},\ref{xisimp},\ref{sigsimp}),
as there the required 3d degree terms have been removed from $\nN$ by
translations of $\ns$ and $\nx$} 
of the metric (\ref{ds2_2}), with $\nN,\nvarsigma,\nxi$ given by (\ref{N_expression},\ref{varsigmasq1},\ref{xisq1}), a limiting procedure, consisting of a coordinate transformation $[\x,\y,\ns,\nx] \rightarrow$
\be
 [\sqrt{2} A^{-1}\x a^{-2},\sqrt{2} A^{-1}\y a^{-2},(m A \ns +\smfrac{1}{6}) a^{4},(m A \nx -\smfrac{1}{6}) a^{4}],
\ee
together with a redefinition of the constants $c_i$, $[c_4,c_5,c_6,c_7,c_8,c_9] \rightarrow$
\be
[\frac{1}{m\sqrt{2}} a^{-6},\frac{\sqrt{2}}{m} a^{-6},\frac{1}{m\sqrt{2}} a^{-6},0,-\smfrac{1}{6} a^8,(\smfrac{1}{54}-m^2 A^2) a^{12}],
\ee
reduces, after performing the limit $a \rightarrow 0$, both cases $e=\pm 1$ of (\ref{ds2_2}) to the vacuum C-metric\cite{EhlersKundt1962,LeviCivita,Weyl1917,GrifPod}, 
\be
\ud s^2 = \frac{1}{A^2 (\nx+\ns)^2} ( - F \ud \x^2 + G \ud \y^2 + \frac{1}{F} {\ud \nx}^2 +\frac{1}{G} \ud \ns^2   ),
\ee
with $F=-1 +\nx^2 -2 A m \nx^3$ and $G=1-\ns^2-2 A m \ns^3$.\\
Whether, in addition, a non-trivial sub-case of the charged C-metric can be obtained
by a limiting procedure (including a singular coordinate transformation, as discussed
in \cite{Paivaetal}) is not clear, since any attempts at removing the $\nx \ns^2-\ns \nx^2$ term from $\nN$
tend to switch off the Maxwell field.\\

From the GHP equations obtained for $\rho,\mu,\pi$ and $\tau$ in \S2 it is clear\cite{McLenVdB1993} that a valence 2 Killing spinor\cite{WalkerPenrose70} exists. There is more: the form of (\ref{ds2_2}) suggests that one should have 
a closer look at the metric when $\nN=k=1$, with $\nvarsigma=\nvarsigma(\ns)$ and $\nxi=\nxi(\nx)$. It is easy to verify that for this metric all $(0,0)$-weighted GHP spin coefficients vanish, while the only non-0
curvature components are $R, \Phi_{11}$ and $\Psi_2$, with $\Psi_2=-\frac{R}{12}$ and
\begin{eqnarray}
e(\nxi \nxi_{,\nx \nx}+\nxi^2_{,\nx}) +\frac{R}{8}-\Phi_{11} &=0,\\
(\nvarsigma \nvarsigma_{,\ns \ns}+\nvarsigma^2_{,\ns}) +\frac{R}{8}+\Phi_{11} &=0,
\end{eqnarray}
showing that this is one of the Killing-Yano spaces studied in \cite{DietzRudiger1}.

\section {Acknowledgment}
All calculations were done using the Maple symbolic algebra system. The properties of the Killing-Yano space, obtained by putting $\nN=1$, were checked with Maple's DifferentialGeometry package\cite{Anderson_Torre}.

\section{Appendix: Ricci, Maxwell and Bianchi equations in the GHP formalism}\label{appendix1}
Below we list some relevant information from the Geroch-Held-Penrose formalism (weights,  commutators and prime operation and Ricci-Maxwell and Bianchi equations) for the special case of vanishing cosmological constant ($\Lambda=0$).
Note that $\Phi_{ij}=\Phi_i \overline{\Phi_j}$. \\

\noindent Weights 
\footnote{Objects $x$ transforming under boosts and rotations as
$x \rightarrow A^{\frac{p+q}{2}}e^{i\frac{p-q}{2}\theta} x$ are called {\em well-weighted of type} $\left(p,q\right)$.}
of the spin-coefficients, the Maxwell and Weyl spinor components and the GHP operators:
\begin{eqnarray}
&\kappa : (3, 1), \nu : (-3, -1), \sigma : (3, -1), \lambda : (-3, 1), \nonumber \\
&\rho : (1, 1), \mu : (-1, -1), \tau : (1, -1), \pi : (-1, 1), \nonumber \\
&\Phi_0 : (2, 0), \Phi_1 : (0, 0), \Phi_2 : (-2, 0), \nonumber \\
&\Psi_0 : (4, 0), \Psi_1 : (2, 0), \Psi_2 : (0, 0), \Psi_3: (-2,0), \Psi_4 : (-4, 0) ,\nonumber \\
&\eth : (1, -1), \etd : (-1,1), \thd : (-1,-1), \tho : (1,1). \nonumber
\end{eqnarray}

\noindent The GHP operators are related to the NP operators by
\begin{eqnarray}
 \tho \eta &= (D - p \epsilon - q \overline{\epsilon}) \eta, \ \thd \eta &= (\Delta -p \gamma-q \overline{\gamma}) \eta , \nonumber \\
 \eth \eta &= (\delta -p \beta -q \overline{\alpha}) \eta, \ \etd \eta &= (\overline{\delta} -p \alpha -q \overline{\beta}) \eta . \label{GHP_NP}
\end{eqnarray}
for any $(p,q)$-weighted scalar $\eta$.\\

\noindent The prime operation is an involution 
with
\begin{eqnarray}
 \kappa' &= -\nu,\sigma'=-\lambda,\rho'=-\mu, \tau'=-\pi,\\
 {\Psi_0}' &= \Psi_4, {\Psi_1}'=\Psi_3, {\Psi_2}'=\Psi_2,\\
 \Phi_0' &= -\Phi_2, \Phi_1'=-\Phi_1.
\end{eqnarray}
and satisfies $\overline{\et}=\etd$.\\

\noindent The GHP commutators acting on $(p,q)$-weighted quantities are given by:
\begin{eqnarray}
\fl \left[ \tho,\tho' \right] &= (\pi+\tauc)\et +(\pic+\tau)\etd +(\kappa\nu-\pi\tau -\Phi_{11}-\Psi_2)p \nonumber \\
\fl &+(\kc\nuc-\pic\tauc -\Phi_{11}-\Psic_2)q,\\
\fl \left[ \et,\etd \right]  &= (\mu-\muc)\tho  +(\rho-\rhoc)\tho' +(\lambda\sigma-\mu\rho-\Phi_{11}+\Psi_2)p \nonumber \\
\fl &-(\overline{\lambda\sigma}-\muc\rhoc-\Phi_{11}+\Psic_2)q,\\
\fl \left[ \tho,\et \right] &= \pic\,\tho -\kappa\tho' +\rhoc\,\et +\sigma\etd+(\kappa\mu-\sigma\pi-\Psi_1)p + (\overline{\kappa\lambda}-\pic\rhoc-\Phi_{01})q .
\end{eqnarray}

\noindent Ricci equations:

\begin{eqnarray}
\tho\rho-\etd\kappa &= \rho^2+\sigma\sigmac-\kc\tau+\kappa\pi+\Phi_{00}, \label{ghp1}\\
\tho\sigma-\et\kappa &= (\rho+\rhoc)\sigma+(\pic-\tau)\kappa+\Psi_0, \label{ghp2}\\
\tho\tau-\tho'\kappa &= (\tau+\pic)\rho+(\tauc+\pi)\sigma+\Phi_{01}+\Psi_1, \label{ghp3}\\
\tho \nu-\tho' \pi  &=  (\pi+\tauc)\mu+(\pic+\tau)\lambda+\Psi_3+\overline{\Phi_1}\Phi_2,\label{ghp6}\\
\et\rho-\etd\sigma &= (\rho-\rhoc)\tau+(\mu-\muc)\kappa+\Phi_{01}-\Psi_1,\label{ghp8}\\
\tho'\sigma-\et\tau &= -\sigma\mu-\lc\rho-\tau^2+\kappa\nuc-\Phi_{02},\label{ghp4d}\\
\tho'\rho-\etd\tau &= -\muc\rho-\lambda\sigma-\tau\tauc+\kappa\nu-\Psi_2. \label{ghp5d}
\end{eqnarray}

\noindent Maxwell equations:
\begin{eqnarray}
\tho \Phi_1-\etd \Phi_0  &= \pi \Phi_0+2\rho\Phi_1-\kappa \Phi_2, \label{max1}\\
\tho \Phi_2-\etd \Phi_1  &= -\lambda \Phi_0+2 \pi \Phi_1+\rho\Phi_2. \label{max2}
\end{eqnarray}

\noindent Bianchi equations:
\begin{eqnarray}
\fl {\etd} \Psi_{{0}}  -{\tho} \Psi_{{1}}
 +{\tho} \Phi_{{01}}  -{\et} \Phi_{
{00}}  &= -\pi\,\Psi_{{0}}-4\,\rho\,\Psi_{{1}}+3\,\kappa\,\Psi_{
{2}}+   \pic    \Phi_{{00}}+2\,
\rhoc    \Phi_{{01}}+2\,\sigma\,\Phi_{{10}} \nonumber \\
\fl &-2\,\kappa\,\Phi_{{11}}-\kc    \Phi_{{02}}, \label{bi1}\\
\fl {\thd}   \Psi_{{0}}    -{\et}   \Psi_{{1}}
    +{\tho}   \Phi_{{02}}    -{\et}   \Phi_{
{01}} &= -\mu\,\Psi_{{0}}-4\,\tau\,\Psi_{{1}}+3\,\sigma\,\Psi_{
{2}}-\lc    \Phi_{{00}}+2\,   \pic    \Phi_{{01}}+2\,\sigma\,\Phi_{{11}}\nonumber \\
\fl &+   \rhoc    \Phi_{{02}}-2\,\kappa\,\Phi_{{12}}, \label{bi2}
\end{eqnarray}

\begin{eqnarray}
\fl &3\,{\etd}   \Psi_{{1}}    -3\,{\tho}   \Psi_{{2}}
    +2\,{\tho}   \Phi_{{11}}    -2\,{\et}
\Phi_{{10}}    +{\etd}   \Phi_{{01}}    -{\thd}
   \Phi_{{00}} = 3\,\lambda\,\Psi_{{0}}-9\,\rho\,\Psi_{{2
}}-6\,\pi\,\Psi_{{1}}+6\,\kappa\,\Psi_{{3}}+ (\muc -2\,\mu )   \Phi_{{00}}\nonumber \\
\fl & \ \  +   2\,(\pi+
   \tauc  )    \Phi_{{01}}+2\,  ( \tau+   \pic  )      \Phi_{{10}}+2\,  ( 2\,
   \rhoc    -\rho )   \Phi_{{11}}
   +2\,\sigma\,\Phi_{{20}
}-\sigmac    \Phi_{{02}}-2\,\kc    \Phi_{{12}}-2\,\kappa\,\Phi_{{21}}, \label{bi3}\\
\fl &3\,{\thd}   \Psi_{{1}}    -3\,{\et}   \Psi_{{2}}
    +2\,{\tho}   \Phi_{{12}}    -2\,{\et}
\Phi_{{11}}    +{\etd}   \Phi_{{02}}    -{\thd}
   \Phi_{{01}}  = 3\,\nu\,\Psi_{{0}}-6\,\mu\,\Psi_{{1}}-9
\,\tau\,\Psi_{{2}}+6\,\sigma\,\Psi_{{3}}-\nuc
    \Phi_{{00}}\nonumber \\
\fl &  \ \ \ +2\, (\muc  -\mu)
    \Phi_{{01}} -2\,\lc    \Phi_{{10}
}+2\,  ( \tau+2 \pic)        \Phi_{{11
}}+   (2\,\pi+\tauc)        \Phi_{{02}} +   2\,(\rhoc    -\rho)    \Phi_{{12
}}+2\,\sigma\,\Phi_{{21}}-2\,\kappa\,\Phi_{{22}}. \label{bi4}
\end{eqnarray}

\end{document}